\documentclass[aps,pre,twocolumn,showpacs]{revtex4}
\usepackage{graphicx,color,epsfig}
\begin{document}
\title{Nonequilibrium transition induced by mass media in a model for social influence}
\author{J. C. Gonz\'alez-Avella$^1$,  M. G. Cosenza$^1$, and K. Tucci$^{1,2}$}
\affiliation{$^1$Centro de
F\'{\i}sica Fundamental, Universidad de Los Andes,
M\'erida~5251, Venezuela \\
$^2$SUMA-CeSiMo, Universidad de Los Andes, M\'erida, Venezuela}
\date{\today}
\begin{abstract}
We study the effect of mass media, modeled as an applied external
field, on a social system based on Axelrod's model for the
dissemination of culture. The numerical simulations show that the
system undergoes a nonequilibrium phase transition between an
ordered phase  (homogeneous culture) specified by the mass media
and a disordered (culturally fragmented) one. The critical
boundary separating these phases is calculated on the parameter
space of the system, given by the intensity of the mass media
influence and the number of options per cultural attribute.
Counterintuitively, mass media can induce cultural diversity when
its intensity is above some threshold value. The nature of the
phase transition changes from continuous to discontinuous at some
critical value of the number of options. 
\end{abstract}
\pacs{89.75.Fb, 87.23.Ge, 05.50.+q}
\maketitle

In recent years, the research on complex systems has extended to
social science in order to understand how collective behaviors
arise in social systems. Several mathematical models, inspired by
analogies with physical systems, have been proposed to describe a
variety of phenomena occurring in social dynamics
\cite{Weidlich,Stauffer,Arrow,MaxiR}. Processes such as
self-organization, cooperation, epidemic spreading, opinion
formation, propagation of information, economic exchanges and
evolution of social structures have been studied by means of
discrete-time, discrete-space dynamical  systems. In this context,
there has been interest in the model introduced
by Axelrod \cite{Axelrod}  to investigate the dissemination of
culture among interacting agents in a social system
\cite{Castellano,Vilone,Maxi1,Maxi2,Shibanai,Maxi3}. From the point of
view of statistical physics, this model is appealing because it
exhibits nontrivial out of equilibrium dynamics, as in other well
studied systems with phase ordering properties \cite{Chate}. Studies on this model
have mainly focused on the collective properties that
result from the interactions between the elements representing endogenous social influences.

In this paper we investigate the effect of external cultural
influences such as mass media on a social system. Our approach is
based on the adaptive interaction dynamics of Axelrod's model for
the dissemination of culture. Agents can interact with their
neighbors in the system and with the mass media according to the
cultural similarities that they share, in each case. The concept
of culture is intended here as a set of individual features or
attributes that are subject to social or external influence.  The
numerical simulations show that, depending on the value of a
parameter that represents the intensity of the mass media
influence and on the number of options available per cultural
attribute, the system displays a phase transition between an
specific ordered phase (a homogeneous culture) imposed by the mass
media and a disordered (culturally fragmented) phase.
Surprisingly, mass media can induce cultural diversity when its
intensity is above some threshold value. The nature of this
transition changes from continuous to discontinuous when the
number of options per cultural feature is increased.

The model consists of $N$ agents as the sites of a square lattice.
The cultural state $c_i$ of agent $i$ is defined as a vector of
$F$ components (cultural features) $c_i=
(\sigma_{i1},\sigma_{i2},\ldots,\sigma_{iF})$. Each $\sigma_{if}$
can take any of the $q$  values in the set $\{ 0, 1, \ldots, q-1
\}$  (cultural traits), initially assigned randomly with a uniform
distribution. There are $q^F$ possible cultural vectors. We define
a mass media cultural message as a vector
$M=(\mu_{1},\mu_{2},\ldots,\mu_{F})$, where $\mu_{f} \in \{ 0, 1,
\ldots, q-1 \}$, that can interact with any of the agents in the
system. We also define a parameter $B \in [0,1]$ that measures the
relative  intensity of the mass media message with respect to the
local interactions, or the probability that  this message has to attract the
attention of the agents in the system. The parameter $B$ represents
enhancing factors of the transmitted message that can be varied
externally, such as its amplitude, frequency, attractiveness,
etc. It is assumed that $B$ is uniform, i.e., the mass media
message reaches all the agents with the same intensity, as a
uniform field. At any given time, we assume that any agent can
either interact with the mass media message or with other agents
in the system. Thus each agent in the network possesses a
probability $B$ of interacting with the message and a probability
$(1-B)$ of interacting with its neighbors.

The discrete-time dynamics of the system subject to the mass media
influence is defined by iterating the following steps:

(1) Select at random an element $i$ in the lattice (active
element) having a cultural state $c_i$. Its attention is drawn to
interact with the message $M$ with probability $B$.

If the attention of element $i$ is drawn to the message, then

(2) Calculate  the cultural overlap (number of shared features)
between the active element and the message
$l(i,M)=\sum_{f=1}^{F}\delta_{\sigma_{if},\mu_{f}}$.

(3) If $0<l(i,M)<F$, the element $i$ and the message interact with
probability $l(i,M)/F$. In case of interaction, choose $h$
randomly such that $\sigma_{ih}\neq \mu_{h}$ and set $\sigma_{ih}
= \mu_{h}$.

If the attention of element $i$ is not caught by the message, then

(4) Select at random a  site $j$ in the nearest neighborhood of site $i$.

(5)  Calculate the cultural overlap
$l(i,j)=\sum_{f=1}^{F}\delta_{\sigma_{if},\sigma_{jf}}$.

(6) If $0<l(i,j)<F$, sites $i$ and $j$ interact with probability
$l(i,j)/F$. In case of interaction, choose $h$ randomly such that
$\sigma_{ih}\neq \sigma_{jh}$ and set $\sigma_{ih} = \sigma_{jh}$.

In this model, the fixed mass media cultural message $M$ can
interact with any element in the system with the same intensity or
probability $B$, but the effect of that interaction may be
different on each element, depending on the specific cultural
overlap between the element and the message. For simplicity, we
are assuming a fixed mass media cultural message acting uniformly
over the system, as for example in a global broadcasting; however
several variations of this condition can be implemented by
extending this basic algorithm. The case $B=0$ corresponds to the
original Axelrod's model.

In any finite network the dynamics settles into an absorbing  or
frozen state, characterized by either $l(i,j)=0$ or $l(i,j)=F$,
$\forall i,j$. Homogeneous or monocultural states correspond to
$l(i,j)=F$,  $\forall i,j$, and obviously there are $q^F$ possible
configurations of this state.  Inhomogeneous or multicultural
states consist of two or more homogeneous domains interconnected
by elements with zero overlap. A domain is a set of contiguous
sites with identical cultural traits. In the absence of external
influences it has been shown that the system reaches ordered,
monocultural states for $q < q_o$ and disordered, multicultural
states for $q > q_o$, where $q_o$ is a critical value that depends
on $F$ \cite{Castellano,Vilone,Maxi1,Maxi2}. This order-disorder
phase transition becomes better defined as  $N$ increases and it
has been argued that it is of first order in two-dimensional systems \cite{Castellano,Maxi1}, while this transition is of second order in
one-dimensional systems \cite{Maxi3}.
The critical value of the number of traits for $F=10$ has been
numerically estimated at $q_o \approx 55$  in two dimensions \cite{Maxi1}.

When the mass media cultural influence is applied to the system,
the order-disorder phase transition persists, but the critical
value  $q_o$ for which the transition takes place decreases as the
intensity of the message $B$ is increased. Figure~1 shows the
spatial configurations of the final states of the system subject
to mass media,  with $F=10$ and $q=35 < q_o$.  In the absence of
external message, i.e.  $B=0$, the system settles into any of the
possible $q^F$ monocultural states. When the intensity of the
message is increased, the system is driven towards a monocultural
state  equal to the state of the mass media, i.e.,  $c_i=M ,\;
\forall i$.  However, there is a critical value of the intensity
$B_c \approx 0.05$  above which the system no longer converges to
the state of the message $M$ but reaches a multicultural state
consisting of an increasing number of domains as $B$ is increased.
Domains having a cultural state equal to the mass media survive
above the threshold intensity $B_c$, but they become smaller in
size as $B$ is increased. Thus, we find the counterintuitive
result that, above some threshold value of intensity, mass media
actually promotes cultural diversity in the system.

\begin{figure}[h]
\includegraphics[width=0.71\linewidth, angle=0]{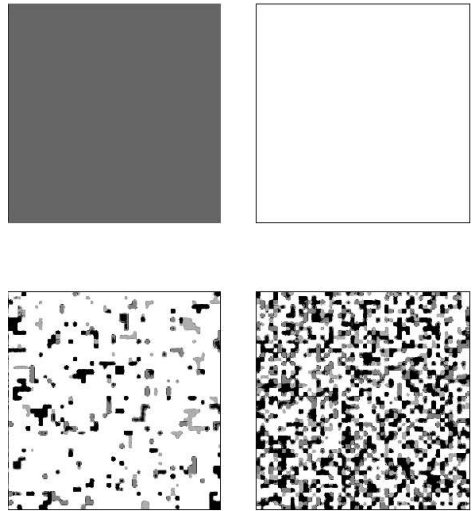}
\caption{Spatial patterns for different values of the intensity
$B$, for $F=10$, $q=35$, and $N= 50 \times 50$. The color code of
the mass media state $M$ is white. Top left: $B=0$; top right:
$B=0.005$; bottom left:  $B=0.1$; bottom right: $B=0.9$.}
\end{figure}

In order to characterize the transition  from the monocultural
state imposed by the message $M$ to a multicultural state when the
intensity $B$ is varied, we consider as an order parameter the
average fraction of cultural domains $g=\langle N_g \rangle /N$,
where $N_g$ is the number of domains  formed in the final state
for a given realization of initial conditions \cite{Note}. Figure~2 shows $g$
as a function of $B$ for different values of the number of traits
$q$. For values of $q>q_o$, the system always reaches a
multicultural state  (with $\langle N_g \rangle  \gg 1$),
independently of the intensity $B$ of the message.  On the other
hand, for each value of $q < q_o$,  we observe a transition at a
critical value $B_c$, from the homogeneous cultural state imposed
by the  message, characterized by values $g\ll 1$, to a
multicultural state for which $g$ increases with $B$. The critical
value $B_c$ decreases with increasing $q$  for $q<q_o$. The
variation of the order parameter $g$ near the value $B_c$ can be
characterized by a critical exponent $\beta(q)$ as the scaling
relation $g\sim (B-B_c)^{\beta(q)}$.  Figure~2 also shows log-log
plots of $g$ vs. $(B-B_c)$ for different values of $q<q_o$. The
exponent $\beta$ as a function of $q$ can be calculated from the
slope of each curve.

\begin{figure}[h]
\includegraphics[width=0.70\linewidth, angle=90]{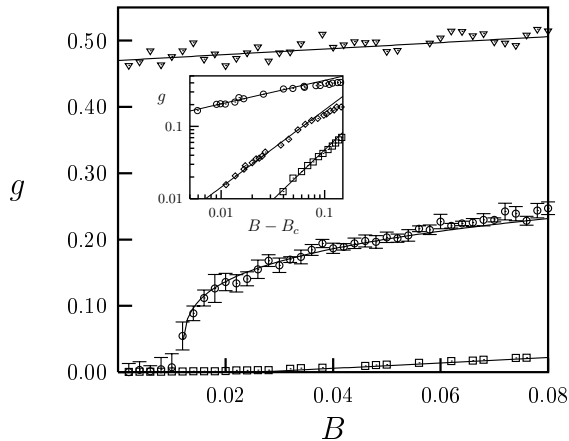}
\caption{Order parameter $g$ as a function of the intensity $B$
for different values of $q$, with $F=10$. Error bars indicate
typical standard deviations about mean values obtained  over $10$
realizations for each value of $B$. The inset shows a log-log plot
of $g$ vs. $(B-B_c)$ for different values of $q$.  Size $N=50
\times 50$. Values of $q$ are $20$  (squares),  $30$  (diamonds),
$45$  (circles), and $60$ (triangles).}
\end{figure}

In Fig.~3 we show the resulting graph of $\beta$ vs. $q$, for $q <
q_o$ corresponding to $F=10$. The dependence of the exponent
$\beta$ with $q$ is well accounted by the linear relation
$\beta(q) \propto (q_o-q)$. As $q$ increases towards the value
$q_o$, the exponent $\beta$ becomes smaller and the corresponding
phase transition from the monocultural state induced by the
message $M$ to multicultural state gets more abrupt. The character
of the phase transition changes from higher order ($\beta > 1$ )
to  second order ($0<\beta<1$) at $q \approx 30$.  At the value
$q=q_o$ for which the exponent $\beta$ vanishes, the character of
the transition  should change from second order to first order.
Figure~3 shows the extrapolation of the straight line  $\beta(q)$
until its intersection with the $q$ axis, predicting a critical
value  $q_o=54$ for the occurrence of a first order transition.

\begin{figure}[h]
\includegraphics[width=0.70\linewidth, angle=90]{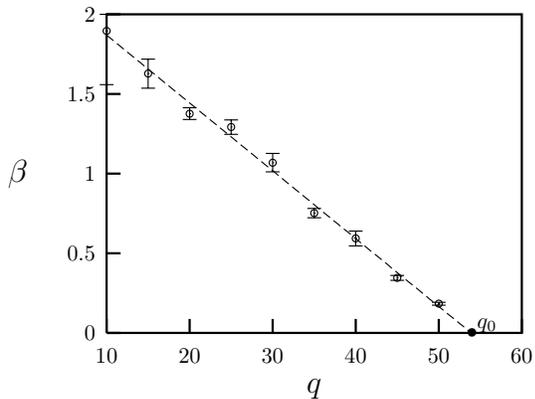}
\caption{Critical exponent $\beta$ for the phase transition
induced by mass media, as a function of $q$.}
\end{figure}

The critical values of the intensity of the message $B_c$ as a
function of $q$, calculated from Fig.~2, are shown in Fig.~4. The
points  in Fig.~4 are well fitted by the relation $B_c \propto
(e^{-k_1 q}-e^{-k_1 q_o})$, which yields $B_c=0$ for $q=q_o=54$ as
indicated in that figure.  Thus, in absence of mass media, the
critical value of the number of traits for the first order phase
transition from cultural homogeneity to a multicultural state
predicted by our model  is $q_o=54$. This agrees with the
approximated critical value obtained numerically in Ref.
\cite{Maxi1} for the order-disorder transition in the original
Axelrod model with $F=10$.  The critical curve $B_c$ vs. $q$ in
Fig.~4 separates the multicultural region from the region where
the monocultural state induced by the mass media occurs on the
space of parameters $(B,q)$. This critical curve reflects the
competition between the  tendency towards the specific  ordering
imposed by the mass media and the disorder present in the initial
configuration of the system, represented by the number of options
per feature $q$. Figure~4 reveals that the ability of the mass
media message to transmit its state on the entire system decreases
as the number of options per cultural feature increases. Above
this threshold curve, the mass media is not longer able to drive
the system towards homogeneity, and a multicultural state sets in.
The above results suggest that the order parameter $g$ behaves,
near the critical boundary, as $g(B,q) \sim \left[ B-k_2\left(
e^{-k_1 q}-e^{-k_1 q_o}\right) \right] ^{k_3(q_o-q)}$, where
$k_1$, $k_2$ and $k_3$ are constants.

\begin{figure}[h]
\includegraphics[width=0.70\linewidth, angle=90]{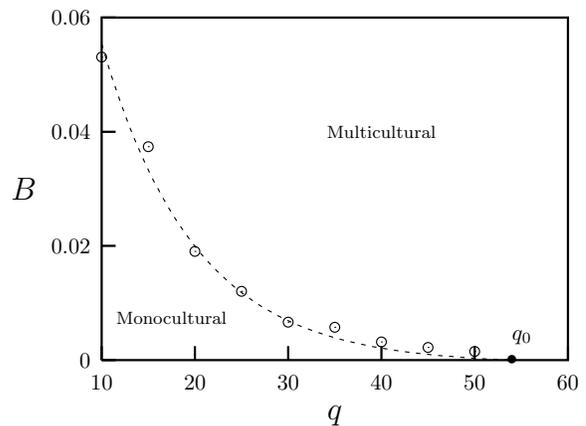}
\caption{Critical boundary $B_c$ vs. $q$ separating  the region of
the monocultural phase induced by mass media from that of
multicultural states, as indicated. The dashed line corresponds to
the fitting of the points. Fixed $F=10$.}
\end{figure}

In the presence of the mass media message, the system reaches
frozen configurations of each phase (homogeneous or multicultural)
much faster than in the original Axelrod model having only local
interactions.  In Fig.~5 we show a log-log plot of the average
time $\tau$ to reach a frozen configuration as a function of the
intensity $B$, for several values of $q<q_o$. For $B<B_c$, the
time $\tau$ decays according o $\tau \sim B^{-0.95}$,
independently of $q$  for $q<q_o$ (where the exponent is
determined from the slopes of the curves in Fig.~5). The times for
convergence to the monocultural state induced by the message  are
much longer than the corresponding times to reach a multicultural
state beyond the critical boundary in Fig.~4. Therefore, for
values of intensity $B<B_c$ the mass media message is able to
transmit its cultural state to all the elements in the system, but
the process takes longer times than the partial convergence to the
message state achieved when $B>B_c$.

\begin{figure}[h]
\includegraphics[width=0.70\linewidth, angle=90]{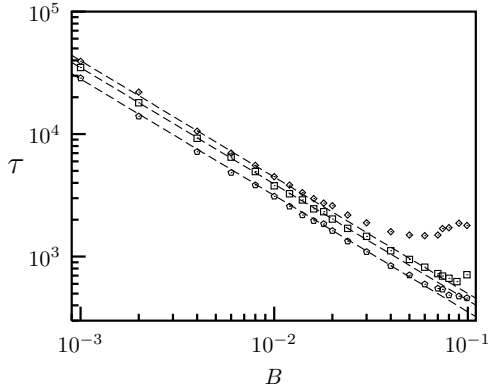}
\caption{Log-log plot of the average transient time to freezing
$\tau$ as a function of $B$, for different values of $q<q_o$. Here
$q=10$ (pentagons), $q=20$ (squares),  $q=30$ (diamonds). 
%Dashed lines correspond to the linear fit for $B < B_c$.
}
\end{figure}

The interaction with the mass media message $M$ accelerates the
convergence of those vectors $c_i$ in the system that possess an
overlap  $0<l(i,M)<F$ towards the cultural state prescribed by
$M$. The dynamics of the interaction with the mass media is
essentially a coarsening process of such domains. For values of
$B$ and $q$ below the critical boundary, this process takes longer
times  and allows enough local interactions to take place and
spread the cultural state of the message,  leading to the growth
and coalescence of domains with the state $M$  into a single
domain of size comparable to that of the whole system. For values
$(B,q)$ above the critical boundary, the process of convergence to
the message and the formation of domains with this cultural vector
occurs faster; this contributes to the early distinction of the
domains with state $M$ from the  other domains and to limit their
growth. Thus, the rapid convergence towards the mass media message
when its intensity is above a threshold value ends up inducing
cultural diversity in the system. The transition between the two
phases in Fig.~4 can also be seen as separating two different
dynamical regimes; one  having a fast decay time for values
$(B,q)$ above the critical boundary and another characterized by a
slow decay for  values $(B,q)$ below that boundary.

In summary, we have presented a model for the influence of mass
media on a social system as an extension of Axelrod's model for
the dissemination of culture. The mass media cultural message has
been assumed as a fixed vector acting uniformly over the system.
We have found the nontrivial result that mass media can actually
induce cultural diversity in conditions for which it was absent,
as for values $q<q_o$ in the original Axelrod's model. We have
verified that this effect persists in lattices with periodic
boundary conditions or with different local geometries.

We have calculated the critical boundary in the space of
parameters $(B,q)$ that separates the monocultural phase induced
by the mass media message from the multicultural phase. These two
phases correspond to different dynamical regimes of the system as
it evolves towards its frozen final state; the first having a slow
decay and the second characterized by a fast decay. The character
of the phase transition changes progressively from continuous to
discontinuous as the number of traits per feature is increased.
The critical value $q_o$ for the onset of the first order phase
transition was predicted from the linear dependence observed in
the critical exponent $\beta$ vs. $q$, as well as from the curve
fitting the critical boundary. The predicted value $q_o$ agrees
with the value estimated previously by numerical simulations in
the original Axelrod's model. The scaling behavior of the order
parameter $g$ characterizing the transition near the critical
boundary was also found numerically. Future extensions of this
basic model should include the consideration of mass media
messages varying in time and/or space, competing messages, noise,
and more complex networks of interactions.

This work was partially supported by grant  C-1285-04-05-A from
Consejo de Desarrollo Cient\'{\i}fico, Human\'{\i}stico y
Tecnol\'ogico of Universidad de Los Andes, Venezuela.

%\pagebreak


\begin{thebibliography}{99}
\bibitem{Note} The order parameter $g$ measures the degree of multiculturality
as defined by Axelrod (number of cultural domains) in \cite{Axelrod}. A related order parameter used in \cite{Castellano,Vilone,Maxi1,Maxi2} is the normalized average size of the largest domain, $\langle S_{\mbox{max}}\rangle/N$. Homogeneity corresponds to
$\langle S_{\mbox{max}}\rangle/N =1$, while complete disorder is given by $\langle S_{\mbox{max}}\rangle/N \ll 1$, for large $N$.
\end{thebibliography}
\end{document}